\documentclass[]{mn2e}
\usepackage{times}

\title[PN around OH\,354.88-0.54]
{Radio observations of the planetary nebula around the OH/IR Star OH\,354.88-0.54 (V1018 Sco)}

\author [M. Cohen et al.]
{Martin Cohen$^{1}$, Jessica M. Chapman$^{2}$, 
Rachel M. Deacon$^{3}$\footnote{Affiliated with the Australia Telescope National Facility, CSIRO}, 
Robert J. Sault$^{2}$, 
\newauthor
Quentin A. Parker$^{4,5}$, Anne J. Green$^{6}$ \\
$^{1}$Radio Astronomy Laboratory, University of California, Berkeley, CA 94720, USA\\
$^{2}$Australia Telescope National Facility, PO Box 76,  Epping, NSW 2121, Australia\\
$^{3}$School of Physics A28, University of Sydney, NSW 2006, Australia\\
$^{4}$Anglo-Australian Observatory, PO Box 296, Epping, NSW 2121, Australia\\
$^{5}$Department of Physics, Macquarie University, Sydney, NSW 2109 Australia\\
$^{6}$Astrophysics Dept., School of Physics, Sydney University, Sydney, NSW 2006, Australia\\ }

\date{ Accepted . Received ; in original form        }

\begin{document}

\maketitle

\begin{abstract}
We present radio observations of the unique, recently formed, planetary nebula (PN) associated 
with a very long-period OH/IR variable star V1018 Sco that is unequivocally still in its asymptotic 
giant branch phase.  Two regions within the optical nebula are clearly detected in nonthermal 
radio continuum emission, with radio spectral indices comparable to those seen in colliding-wind 
Wolf-Rayet binaries.  We suggest that these represent shocked interactions between the
hot, fast stellar wind and the cold nebular shell that represents the PN's slow wind moving
away from the central star.  This same interface produces both synchrotron radio continuum
and the optical PN emission.  The fast wind is neither spherical in geometry nor aligned with
any obvious optical or radio axis.  We also report the detection of transient H$_2$O maser 
emission in this nebula.
\end{abstract}


\section{INTRODUCTION}
Cohen, Parker \& Chapman (2005: hereafter CPC) have reported the discovery of a very faint 
planetary nebula (PN) centred on the long-period pulsating OH/IR star OH\,354.88-0.54, also
known as V1018~Sco.  The PN was found during examination of survey field HA630 of the
Southern H$\alpha$ Survey (SHS: Parker et al. 2005a) by QAP, as part of a systematic search for
new Galactic Plane PNe (e.g. Parker et al. 2005b, 2006). 
Its OH characteristics mark this star as an intermediate mass asymptotic 
giant branch (AGB) star, with initial mass $\geq4M_{\odot}$. After rejecting the possibility
that the PN arises from shock interaction with the surrounding external medium, or that the
star is a ``born-again" (e.g., Iben 1984) object that has
undergone a very late thermal pulse (e.g., Bl\"{o}cker \& Sch\"{o}nberner 1997) as it began to 
evolve along its cooling track, CPC speculated that this object represents a phase of PN evolution 
that has been formerly unobserved.  They concluded that this star's fast wind has very
recently turned on and it is the interaction between this wind and the previously shed outer 
layers of the star that produces the ionization that reveals the optical PN.

To explore this intriguing phenomenon further we have undertaken a series of radio continuum and water maser
observations with the Australia Telescope Compact Array (ATCA), originally with the objective of mapping out
the spatial distribution of ionized material within the PN. Pre-existing radio data are discussed 
in \S2.  \S3 describes the new ATCA continuum and H$_2$O maser observations, while \S4 shows an
enhanced optical image of the PN.  The discussion follows in \S5, with our conclusions in \S6.
 
\section{Previous radio continuum obervations}
The best estimate for the position of the star itself is: 17$^h$ 35$^m$ 02.73$^s$$\pm$0.02$^s$,
$-$33$^\circ$ 33$^\prime$ 29.41$^{\prime\prime}$$\pm$0.24$^{\prime\prime}$ (CPC).  
Neither the PMN or NVSS surveys includes any continuum source within several arcmin of
OH\,354.88-0.5, while the VLA survey of the inner plane at 5~GHz by Becker et
al. (1994) is limited to $|$b$|$$\leq$0.4$^\circ$.  The 1.4-GHz continuum survey of small 
diameter sources by Zoonematkermani et al. (1990) does include a bright, compact radio 
object (2.4$^{\prime\prime}$ in size) within 7$^{\prime\prime}$ of the maser position. 
However, this survey comprises two 3-MHz wide bands centred at 1441.5 and 1611.7~MHz, so 
the higher frequency band is contaminated by the strong 1612-MHz maser emission.
This problem was recognized by White, Becker \& Helfand (2005),who give an updated position
at 20~cm of 17$^h$ 35$^m$ 02.590$^s$ $-$33$^\circ$ 33$^\prime$ 31.18$^{\prime\prime}$ (J2000)
and warn of this contamination of their band-averaged flux densities.
Zijlstra et al. (1989) quote an upper limit of 1.4 mJy at 2~cm from a VLA observation in September 1986.

Consequently, the only radio survey to have correctly detected and assessed the continuum from
OH\,354.88-0.5 is the first Molonglo Galactic Plane Survey (MGPS1: Green {\it et al.} 1999), at
843~MHz.  Extracting the individual MGPS1 images, from which the final survey images were
constructed, shows a source at the maser position.  Fitting a 2D-Gaussian to the object 
indicates an unresolved source with an 843-MHz flux density of 10.8$\pm$0.6~mJy, located at
17$^h$ 35$^m$ 01.52$^s$ $-$33$^\circ$ 33$^\prime$ 36.90$^{\prime\prime}$ (J2000)
The beam size of the Molongolo Observatory Synthesis Telescope at this 
location is 75$^{\prime\prime}$ (N-S) by 43$^{\prime\prime}$ (E-W) so the radio
emission could occupy the full extent of the H$\alpha$ ring (roughly a 39$^{\prime\prime}$
circle as shown by the images in \S4) yet still be unresolved in MGPS1 images.  Therefore, 
observations with the ATCA were planned.

\section{New radio observations}
Short test observations were carried out with the ATCA in the continuum at 3 and
6~cm in December 2003 to determine whether emission could be detected.
The ATCA has six 22-m diameter antennas with five of the antennas
located on an east-west track of length 3~km. The sixth antenna lies 3~km
further west giving a maximum baseline of 6~km.  
Further continuum observations were made at 3, 6, 13 and 20~cm in
2004$-$2005.  Only OH and SiO masers were known for OH\,354.88-0.54.
Therefore, we investigated the H$_2$O maser line at 22235~MHz in
November 2004 with a short observation that yielded a solid detection.
We later followed up with a 12-h observation in June 2005 with the
intention of imaging the water maser emission and to search for
continuum emission at 22235~MHz.

Table~1 summarizes our new continuum observations. We refer to the
observing bands as 1.35, 3, 6, 13 and 20~cm.  The actual frequencies
used were: 22235, 8640, 4800, 2368, and 1384~MHz.  We used a phase
centre of 17$^h$ 35$^m$ 02.73$^s$ $-$33$^\circ$ 33$^\prime$
19.71$^{\prime\prime}$ (J2000) ~for the continuum observations, and
17$^h$ 35$^m$ 02.73$^s$ $-$33$^\circ$ 33$^\prime$ 29.41$^{\prime\prime}$
for the maser observations, corresponding to the stellar position.
To correct for atmospheric amplitude and phase variations, the
observations of the source were interleaved with short observations of a
secondary calibrator source (1729-37 for 3, 6, 13 and 20~cm, and
1710-269 for 1.35~cm). The flux density scale and bandpass calibrations
were applied from observations of the source 1934-638 which is the
primary flux calibrator for the ATCA.  The resolutions achieved are
given in column(4) ot Table~2.

\begin{table*}
\caption{Log of ATCA observations.}
\begin{tabular}{llccclll}
Date& Project& Config& Time& Bands& Cont/line& Bandwidth& No.channels\\
& & & & cm& & MHz& \\
\hline
08 Jan 2004& CX053& 6A& 6h& 3,6& Continuum& 128& 32\\
14 Nov 2004& C1339& 750C& $\sim$45min& 1.35& H$_2$O maser& 32& 512\\
21 Dec 2004& C1339& 1.5D& 12h& 3,6& Continuum& 128& 32\\
24 March 2005& C1339& 6A& 12h& 13,20& Continuum& 128& 32\\
21 June 2005& C1339& 6B& $\sim$12h& 1.35& H$_2$O maser& 16& 512\\
\end{tabular}
\end{table*}

\begin{figure}
\vspace{8cm}
\includegraphics{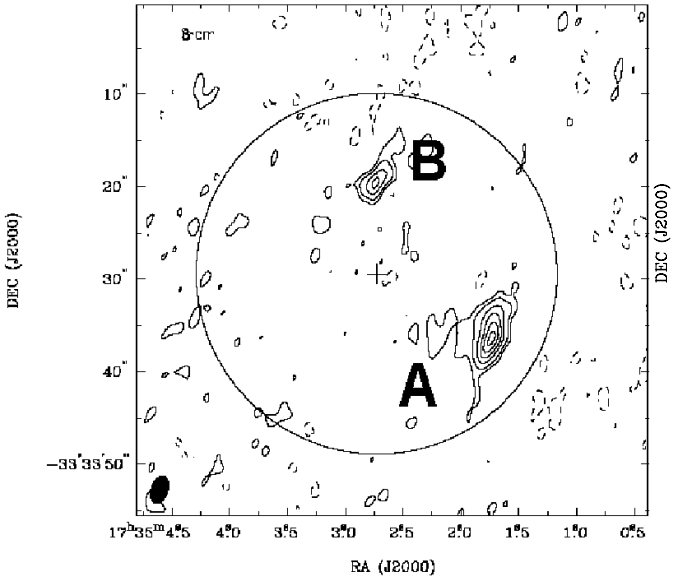}
\caption{3-cm ATCA continuum image of OH\,354.88-0.54. The large
circle indicates the approximate optical extent of the
PN. The ring is 39$^{\prime\prime}$ in diameter, centred on the
stellar position which is marked by the small cross.  Radio contours
plotted correspond to -4, -2, 2, 4, 8, 12, 16, 20~$\sigma$.  1$\sigma$
is 0.04~mJy~beam$^{-1}$.  The beam size is shown in the bottom left
corner.}
\label{3cm}
\end{figure}

\begin{figure}
\vspace{8cm}
\includegraphics{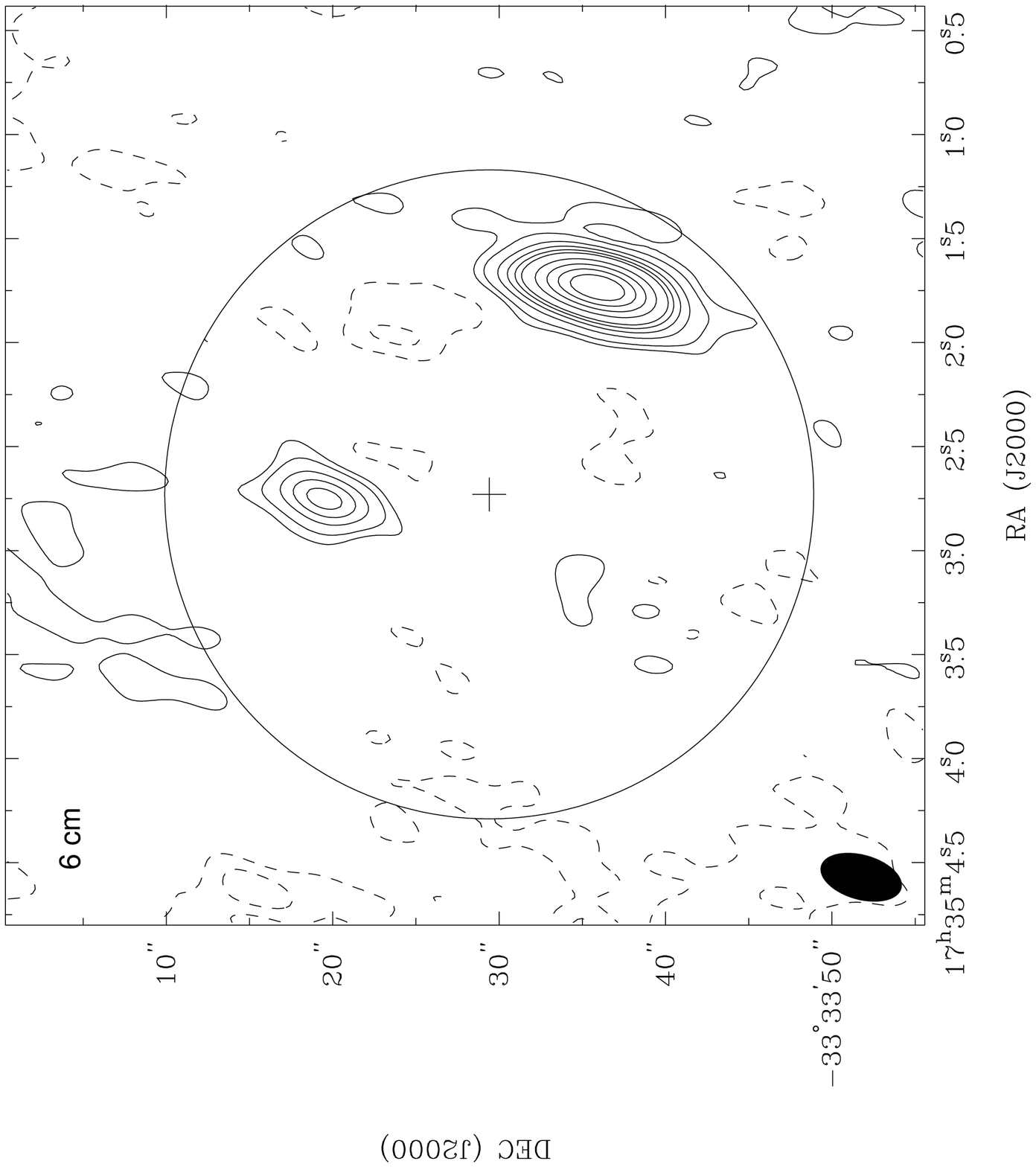}
\caption{6-cm ATCA continuum image of OH\,354.88-0.54 as for Fig.~\ref{3cm}. Radio contours 
plotted correspond to -4, -2, 2, 4, 8, 12, 16, 20, 30, 40, 50, 60~$\sigma$.
1$\sigma$ is 0.03~mJy~beam$^{-1}$.}
\label{6cm}
\end{figure}

\begin{figure}
\vspace{8cm}
\includegraphics{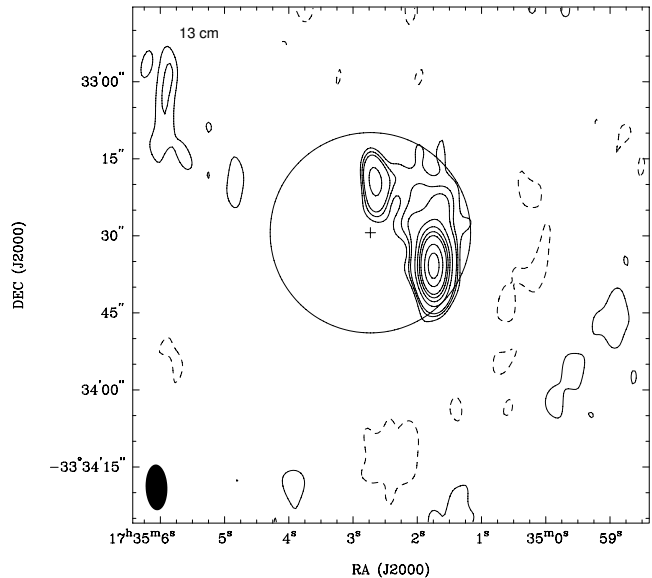}
\caption{13-cm ATCA continuum image of OH\,354.88-0.54 as for Fig.~\ref{3cm}.
Radio contours have been chosen to emphasize a possible weak structure
linking the two continuum components and correspond to -1.5, 1.5, 2, 3,
6, 8, 12, 14, 21, 30, 35 and 40~mJy~beam$^{-1}$. 1$\sigma$ is 0.14~mJy~beam$^{-1}$.}
\label{13cm}
\end{figure}

\begin{figure}
\vspace{8cm}
\includegraphics{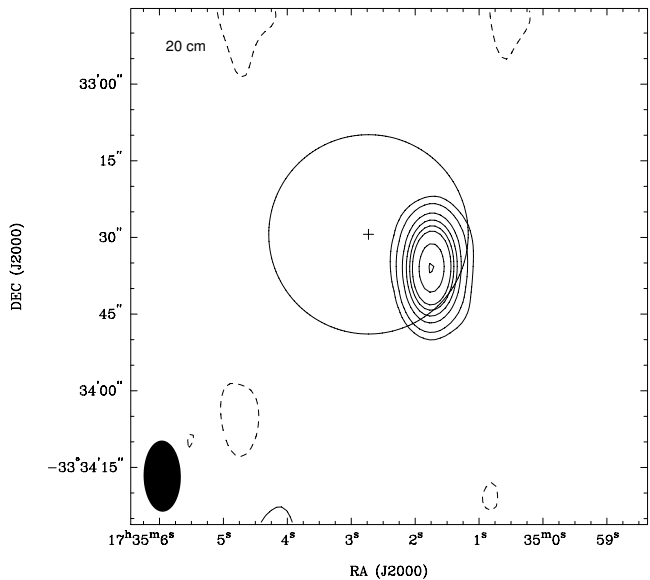}
\caption{20-cm ATCA continuum image of OH\,354.88-0.54 as for Fig.~\ref{3cm}.
Radio contours plotted correspond to -2, 2, 4, 8, 12, 16, 20$\sigma$. 
1$\sigma$ is 0.18~mJy~beam$^{-1}$. }
\label{20cm}
\end{figure}

The 3, 6, 13 and 20-cm continuum images were made using natural weighting.
All 3-cm data were used because there were no indications of broad scale 
Galactic emission.  We cut out 6-cm data with uv-distances $<$2~k$\lambda$ 
to eliminate a low-level extended wash of Galactic emission
and this radically improved the imagery.  For the 13 and 20-cm data we 
excluded data with uv-distances $<$3~k$\lambda$.  Figures~\ref{3cm},
\ref{6cm}, \ref{13cm} and \ref{20cm} show the images with the position of 
the star and the approximate location of the outer optical edge of the PN also plotted.

Two obvious sources are detected within the PN at 3, 6 and 13~cm. The
south western source (source A) is the stronger and is partially
resolved with the resolution of the ATCA. The source to the north of
the stellar position (source B) is unresolved and significantly
weaker, with no detection at 20~cm.

Table~2 presents the flux densities, positions, sizes and offsets from
the star, for source A and source B at each wavelength. For source A,
the size was estimated for each waveband by fitting a Gaussian and then
deconvolving to remove the ATCA restoring beam. For source B, the
Gaussian fitting was consistent in all cases with a point source.

A description of Table~2 follows: col.(1)
gives the wavelength in cm; col.(2) the source; col.(3) the integrated
flux density and its uncertainty; col.(4) the semiaxes and p.a. of the
major axis for the beams used to deconvolve the images; col.(5) the same parameters 
as in col.(4) for the deconvolved sources; col.(6) the best-fit
positions to the Gaussians; and col.(7) the modulus of the vector
distance between each source and the central star.  Our upper 
limits are 5$\sigma$ values.

\begin{figure}
\vspace{7cm}
\includegraphics{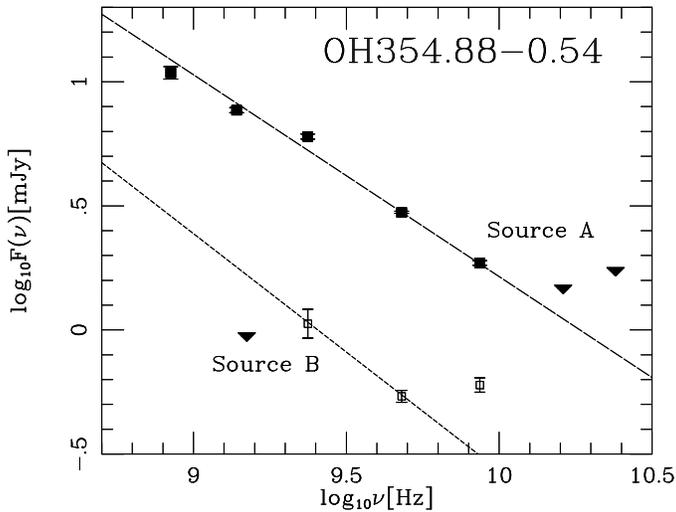}
\caption{Spectral energy distributions of sources A and B fitted by power laws.  Filled 
triangles indicate upper limits at 2~cm from the VLA and 1.3~cm from our data for A, and 
at 20~cm, from our data, for source B.}
\label{sed}
\end{figure}

Source A has a definite nonthermal spectrum with the detected flux density
falling with increasing frequency. The 20-cm detection of 7.7 mJy is
comparable to the 10.8$\pm$0.6~mJy detection of the MGPS1 object (Fig.~\ref{sed}).
Therefore, we associate A with the 843-MHz (35-cm) source.  In Figure 5,
inverse-variance weighted least squares fits to power laws are plotted
for sources A and B. Both sources have strongly nonthermal spectra
(F$_{\nu}$~$\propto$~$\nu^{index}$) with indices of $-$0.81$\pm$0.01 and
$-$0.95$\pm$0.11 respectively, probably due to synchroton emission.
Fitting just the four ATCA data points for source A still yields a
slope of $-$0.81$\pm$0.01, indistinguishable from the power law fit
including the MGPS1 datum, vindicating our identification of source A
as the 843-MHz source.  The power law fitted to A is also consistent
with the VLA 2-cm upper limit obtained by Zijlstra et al. (1989) and
our own limit on continuum at 1.35~cm (see below).

Source B is much weaker than A, but the slope of the radio continuum
emission is defined by robust detections at 6 and 13~cm. Source B
appears to suffer substantial absorption at 20~cm (Fig.~\ref{sed})
which we interpret as free-free absorption.  This absorption is likely
to occur within the volume of the PN suggesting that source A is seen
on the front of the nebula while source B is seen at the back.

The 3-cm flux density of 0.60~mJy detected for source B is somewhat
higher than that expected for synchroton emission alone (0.31~mJy). 
(We did not include this point in the derivation of the spectral
index for B because of the high significance of the deviation from the
slope.)  We interpret this detection as indicating the likely 
presence of some additional weak thermal radio emission (\S5).

In Figure 3, a weak structure appears to provide a link between
sources A and B at 13~cm (Fig.~\ref{3cm}). However, this is only a
2$\sigma$ detection and is not confirmed. There is no evidence for
diffuse radio continuum emission filling the entire PN.

\begin{table*}
\caption{Summary of best-fit continuum results to single Gaussians for sources A and B.}
\begin{tabular}{lllllll}
Band& Source& Total flux$\pm$rms& Restoring beam& Deconvolved& Best-fit& Offset\\
cm&        &          mJy& semi-axes& size& Position& from star\\  
   &        &            & (arcsec)(deg)&  (arcsec)(deg)& & arcsec\\
\hline
1.35& ...& $<$1.65& 1.1$\times$0.6& & No continuum detected\\
3& A& 1.86$\pm$0.04& 3.0$\times$1.83(-20.9)& 4.48$\times$1.39(-12.8)& 17:35:01.74 $-$33:33:36.0& 10.5$\pm$1.4\\
6& A& 2.98$\pm$0.03& 5.1$\times$2.7(-15.7)& 4.4$\times$0.84(-12.5)& 17:35:01.74   $-$33:33:35.9& 10.5$\pm$1.4\\
13& A& 6.02$\pm$0.14& 9.3$\times$4.5(2.3)& 6.0$\times$1.9(-1.6)& 17:35:01.75 $-$33:33:35.6&10.2$\pm$1.4\\
20& A& 7.68$\pm$0.18& 13.9$\times$7.3(0.9)& 2.8$\times$0.65(-41)& 17:35:01.74    $-$33:33:36.0&10.5$\pm$1.4\\
3& B& 0.60$\pm$0.04& 3.0$\times$1.83(-20.9)& Point source&      17:35:02.73 $-$33:33:19.3& 10.1$\pm$1.4\\
6& B& 0.54$\pm$0.03& 5.1$\times$2.7(-15.7)& Point source& 17:35:02.75   $-$33:33:19.5& 10.0$\pm$1.4\\
13& B& 1.06$\pm$0.14& 9.3$\times$4.5(2.3)& Point source& 17:35:02.64    $-$33:33:19.7& 9.7$\pm$1.4\\
20& B& $<$0.90& 13.9$\times$7.3(0.9)& Point source& No continuum detected\\
\end{tabular}
\end{table*}

The projected mean distances from the stellar position to the peaks of sources A and B are the
same, within the uncertainties: 10.4$\pm$0.7$^{\prime\prime}$ (A) and 9.9$\pm$0.8$^{\prime\prime}$ (B).

\begin{figure}
\vspace{7cm}
\includegraphics{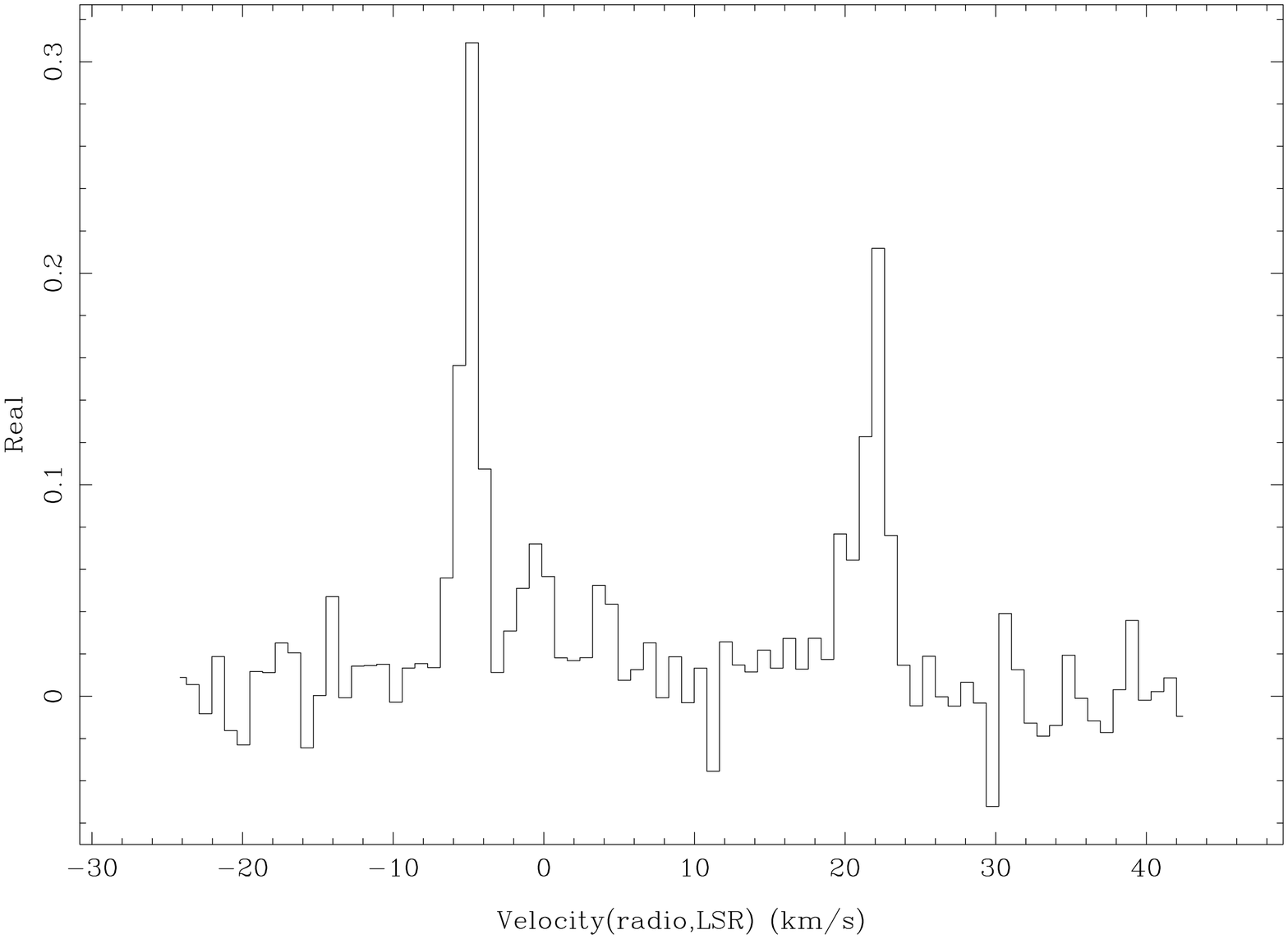}
\caption{Spectrum of H$_2$O maser emission detected from the PN.}
\label{water}
\end{figure}

Figure~\ref{water} represents the discovery of H$_2$O maser emission
from OH\,354.88-0.54, from a 40-minute observation taken in November
2004.  Two peaks are seen at velocities (LSR) of $-$4.8 and
+22.2~km~s$^{-1}$, significantly displaced by 1-2~km~s$^{-1}$ from the corresponding
OH maser peaks ($-$5.5 and +24.3 km~s$^{-1}$, respectively: CPC).  The
blue peak is the stronger with a flux density of 0.31~Jy; the red peak
has a flux density of 0.21~Jy.  From these limited data we were unable 
to measure the maser positions.  We can infer only that we observed two
blobs of water maser emission, but we cannot interpret the peak velocity
differences in terms of relative locations of the masing species.  
It is also noteworthy that the water maser profile is significantly above
zero in the centre, similar to the OH profile (CPC). 

Despite the 12-h track obtained during the June 2005 ATCA observation, we
had a non-detection of any water maser emission with a 5$\sigma$ limit
of $\sim$40~mJy. Extrapolating from Figure~3 of CPC we estimate that
we detected water masers at an OH maser phase of 0.57 but failed at a
phase of 0.72.  This interval is such a small fraction of a period
that the source must be highly variable and the water masers clearly
do not follow the very regular OH maser variations.  The range of
variation in H$_2$O maser intensity from the 2004 to the 2005
observation was about a factor of 8.  We looked for radio continuum
emission at 1.35~cm using the spectral line data but achieved no
detection, with a 5$\sigma$ upper limit of 1.65~mJy.

\section{Optical imagery}
\begin{figure}
\vspace{8cm}
\includegraphics{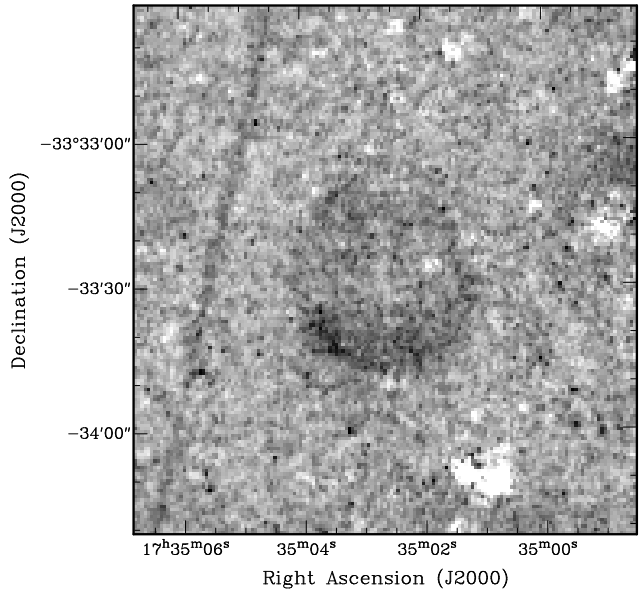}
\caption{H$\alpha$/SR quotient image of the PN with variable PSF-matching before dividing
the H$\alpha$ and SR data.  The linear feature east of the PN is the track of a satellite.}
\label{quotpsf}
\end{figure}

\begin{figure}
\vspace{8cm}
\includegraphics{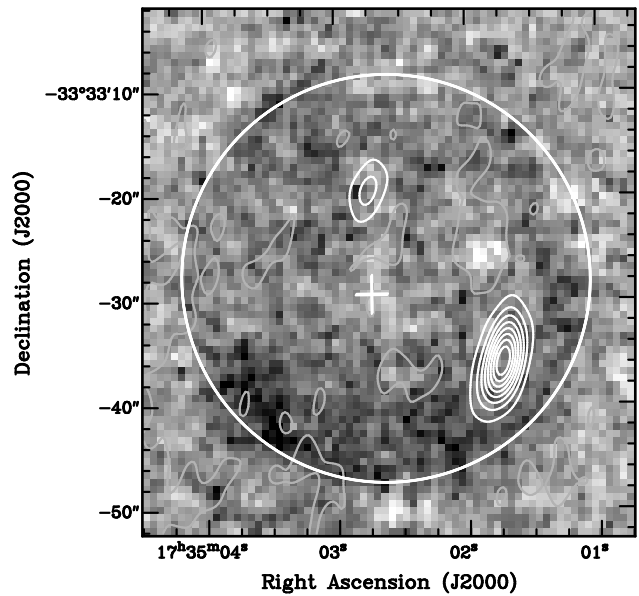}
\caption{An enlarged version of Fig.~\ref{quotpsf} overlaid by white contours of 6-cm radio 
continuum emission corresponding to 0.3, 0.5, 0.8, 1.1, 1.4, 1.6, 1.9, 
2.2~mJy~beam$^{-1}$. White circle marks the approximate outer boundary of the PN; cross marks 
the position of the star and OH masers.}
\label{optrad}
\end{figure}

Figure~\ref{quotpsf} represents a quotient map obtained directly from 
SuperCOSMOS data for the PN by dividing the H$\alpha$ pixel data  by the 
corresponding data from the short-red (SR) exposure.  However, it also involves
use of the Bond et al. (2002) technique of variable point-spread-function (PSF) 
matching between the two images prior to division.  This enhances the faint 
nebula by effectively removing most of the emission from point sources in 
the field.  The figure shows that the outer boundary of the PN is quite
sharply defined around most of its circumference.  Emission can clearly be seen
above the sky over most of its projected area although the brightening of the
south-eastern region is particularly striking.  There is also the suggestion of
a northward extension although this is very faint and merges imperceptibly with 
the interstellar medium (ISM).  The north-south extent could be as much as 20\% greater
than the east-west diameter in this newly processed image.  Fig.~\ref{optrad} 
shows the relationship between sources A and B and the PN in the plane of the sky.

An old UKST H$\alpha$ plate which includes the PN region exists in the archive.
This exposure (HA2307) was taken in 1976 for John Meaburn on fast but
coarse-grained O-9804 emulsion with a mosaic H$\alpha$ filter. The enhanced
south-east section of the PN is clearly visible on this old exposure but the remaining   
boundary is hard to discern. This, coupled with the poorer resolution of this 
exposure, make it essentialy impossible to measure whether there has been 
any expansion of the nebula ring over the intervening 20-plus years.

\section{Discussion}
Nonthermal emission from early-type stars was found in Wolf-Rayet
stars (WRs) by Abbott, Bieging, \& Churchwell 1984; Abbott et
al. 1986; Bieging et al. 1989).  Its origin lies in the acceleration,
by strong shocks, of a small population of relativistic electrons
(White 1985).  In WR binary systems these shocks arise from the
collision of the separate winds of the two components. It is less
certain whether the phenomenon occurs in single WR stars. The
paradigm applied to OH\,354.88-0.54 involves a single star but two
stellar winds from different phases of evolution.  Absorption of
synchrotron emission has been observed in colliding winds from WRs by
Chapman et al. (1999), who attributed this to thermal absorption by
gas along the line-of-sight through the ionized wind.
The observed continuum energy distribution for WR\,48
(Chapman et al.  1999, Fig.~11) reveals nonthermal emission between 3
and 13~cm but only an upper limit at 20~cm, implying a pronounced
turnover longward of $\sim$13~cm.  These authors separated the thermal
and nonthermal contributions to the spectrum of WR\,48 and found the
nonthermal component to have a slope of $-$0.83.  This slope matches
that for OH\,354.88-0.54 sources A and B (within the errors), strongly suggesting that we
may be viewing nonthermal emission in this PN that is also caused by
the collision of two winds, the low density, high velocity PN fast
wind and the high density, slowly moving original PN ejecta.
Synchrotron self-absorption is unimportant in sources with brightness
temperature below 10$^{10}$~K (Condon 1992) but thermal absorption
within the PN is still relevant and could cause the abrupt turn-down
in the spectrum of source B.  Locating source A near the front side of
the PN, as we view it, would account for the lack of absorption in its
radio spectrum.

\subsection{Estimates of electron density}
The PSF-subtracted image of the PN (Fig.~\ref{quotpsf}) indicates that the H$\alpha$ brightness 
varies significantly across the PN suggesting that the electron density also varies.  
It is also possible that the southern rim of the PN is brightest because the PN's slow wind 
suffers a shock at the interface with the ISM.  Source A lies within the 
western portion of the brightest (southern) region of H$\alpha$ emission, for which 
CPC determined N$_e$~$\sim$4000~cm$^{-3}$, whereas B appears in an area of much lower H$\alpha$
brightness to the north (Fig.~\ref{quotpsf}).  We can use the turnover in source B's radio spectrum to
estimate the mean electron density through the nebula in its direction.  The thermal optical depth 
through the nebula may be calculated following Mezger \& Henderson (1967, eqn (5)), and 
rearranged into an expression for the turnover frequency (in GHz), 
$\nu_T$$\approx$0.3(T$_{e}^{-1.35}$N$_{e}^2$L)$^{0.5}$ 
(Lang 1999, eqn.1.224), where L is the path length through the PN in pc (we adopt the diameter of the PN, 
1.9$\times$10$^{18}$~cm, 0.6~pc, for L: note that CPC's Table~1 gave the diameter as 0.3~pc; this was in fact 
the radius). The turnover wavelength must lie beyond 13~cm but be less than 20~cm, so $\nu_T$ must 
lie between 1.5 and 2.3~GHz. Taking T$_e$$\sim$10$^4$~K, we conclude that the mean nebular N$_e$ 
along the line-of-sight toward B is between 3200 and 4900~cm$^{-3}$, in excellent agreement with
the density derived from the [S{\sc ii}] lines (4000~cm$^{-3}$: CPC).

\subsection{Possible free-free emission in source B}
We also consider whether the apparent excess of 3-cm emission in B,
above the nonthermal slope, might be due to free-free emission, as
seen in WR stars.  Firstly we have separately analyzed the two 3-cm data
sets to investigate the reality of this excess by fitting a point source 
to the calibrated u-v data.  This should give a more reliable flux estimate 
than fitting in the image plane, if there were no confusion and the
emission were unresolved.  We find 0.61$\pm$0.03~mJy (Jan 2004) and
0.59$\pm$0.03~mJy (Dec 2004), where the errors are the rms residuals from 
the fitting process alone.  Our measurement of the same value at two epochs 
separated by nearly a year gives strong support for the reality of the
measured flux density.  Extrapolating source B's nonthermal spectrum to 3~cm
implies a level of 0.31~mJy. Therefore, we conclude that B shows a real excess
at 3~cm of 0.29~mJy.

To estimate the expected radio free-free level we have applied an
absolute calibration to the H$\alpha$ image of the nebula according
the method described by Pierce (2006), in which a scaled continuum
(SR) image is subtracted from the H$\alpha$ image, and a factor is
applied to relate the SHS difference image to the spatially much
lower-resolution, but absolutely calibrated, Southern H$\alpha$ Sky
Survey Atlas (SHASSA: Gaustad et al. 2001).  For this PN, the SR scale
is 0.99 and the conversion factor in this field is
14.9~counts~pixel$^{-1}$~per Rayleigh.  Integrating the entire PN
spatially, subtracting background sky (assessed as the average of that
measured in four different locations to provide a robust background)
and three point sources interior to, or on the rim of, the nebula, and
after correction for the 36\% contribution of the red [N{\sc ii}] line
to the emission in the bandpass of the SHS H$\alpha$ filter, we derive
3.6~Rayleighs in H$\alpha$ alone.  This is equivalent to
2.4$\times$10$^{-14}$~erg~cm$^{-2}$~s$^{-1}$.  We now require the total 
extinction to the PN.  Based on Fitzgerald (1968), Lucke (1978) and 
especially on Neckel \& Klare (1980), we find A$_V$$\approx$3.5 in this 
direction, corresponding to A$_{H\alpha}$ $\approx$2.8.  Thus the
intrinsic H$\alpha$ flux is
3.3$\times$10$^{-13}$~erg~cm$^{-2}$~s$^{-1}$, and the corresponding
H$\beta$ flux would be 1.1$\times$10$^{-13}$~erg~cm$^{-2}$~s$^{-1}$.

This is convertible into the radio free-free flux density using Condon
(1992: eqns. (3) and (4a)).  The estimated free-free emission at 3~cm,
for the entire nebula, is 0.33~mJy and this is comparable to the excess 
emission detected at 3~cm.  Source B was unresolved and thus only occupies 
a small part (less than about 2\%) of the nebular area.  For a 
uniform distribution the predicted free-free emission from source B would 
be less than 0.01~mJy.  However, one would expect the radio thermal emission 
to follow the H$\alpha$ brightness distribution and, therefore, to be patchy. 
There is precedent for this in the Luminous Blue Variable HD~168625 
(Leitherer et al. 1995), where the radio continuum shell closely mimics the
morphology of the partially filled H$\alpha$ shell.
If the free-free emission were highly clumped toward source B, then a larger
fraction of the total thermal emission might be concentrated there.
Alternatively, if our line-of-sight to source B through the PN suffered 
significant internal extinction then the estimated intrinsic H$\alpha$ emission 
would also increase.  Consequently, local overdensities of ionized gas could be 
offset by internal nebular extinction due to dusty clumps.  For example, if 20\% 
of the total observed free-free emission were located within source B (an
overdensity greater than a factor of 10), and there were an extra extinction 
(A$_V$) of 2.1 magnitudes (A$_{H\alpha}$ of 1.7 mag) within the nebula, 
then radio free-free emission of 0.3~mJy would still be observed.  (In this case 
the total PN free-free emission would be 1.65~mJy, boosting the intrinsic Balmer line fluxes
by $\times$5, requiring a line-of-sight A$_{H\alpha}$ of 4.6, equivalent to an 
A$_V$ of 5.6~mag, of which 3.5~mag arise from the ISM and 2.1~mag within the PN.

The upper limit on B at 20~cm also excludes a cut-off in the electron
energy spectrum because there is no intermediate regime with
S$_\nu$$\propto$$\nu^{1/3}$ between the optically thin and thick
spectral indices (Ginzburg \& Syrovatskii 1965). The fact that A and B
have almost the same nonthermal slope strengthens the case for either a
common volume in which we observe this nonthermal emission, or two
separated volumes in which very similar strong shocks generate
relativistic electrons.  The fast wind, therefore, must encounter
essentially the same physical conditions at the two locations in which
it impacts the nebular ejecta.

\subsection{Background source contamination}
\begin{figure}
\vspace{8.5cm}
\includegraphics{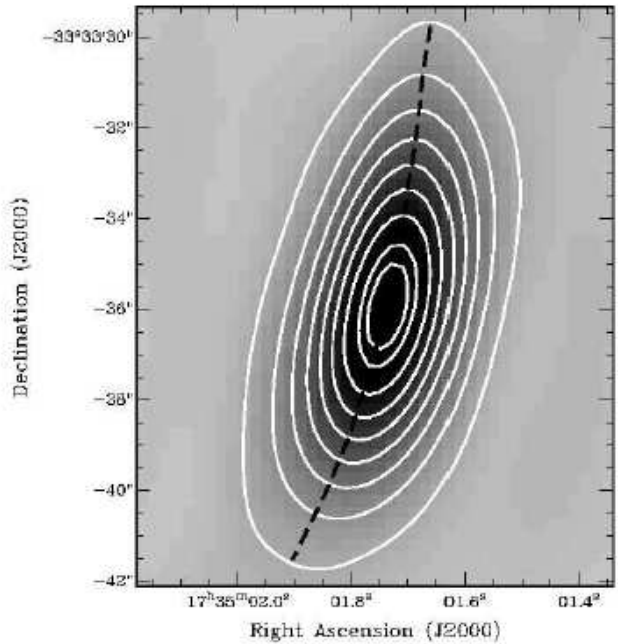} 
\caption{6-cm continuum emission of source A shown both as a greyscale image and overlaid 
by its own contours in white, for levels of 0.2, 0.4, 0.7, 0.9, 1.2, 1.4, 1.6, 1.85, and 
2.0~mJy~beam$^{-1}$.  The black dashes mark the locus of the ends of the major axes of the 
white contours.  Note the distinct curvature of this locus.}
\label{curve}
\end{figure}

Both sources A and B are located within the ring and no other strong background objects are apparent.
One can ask whether sources A and B are truly associated with the PN or might represent
a double radio source viewed by chance along the line-of-sight.  Inspection of our observed images 
covering the ATCA primary beams at 20~cm ($\sim$1000~arcmin$^2$) and 6~cm ($\sim$80~arcmin$^2$) suggests 
that 7 and 3 sources, respectively, are recognizable above the noise, at all flux 
densities, and none appear double.  Calculating the formal extragalactic confusion from
Log~N$-$Log~S relationships at 20 and 6~cm (Ledden et al. 1980; Bridle 1989) one would expect to detect 
15 radio sources within the ATCA primary beam at 20~cm with a flux density greater than the sum 
of A and B, but less than one similarly within the 6-cm primary beam.  The probabilities
of even a single background source as bright as A appearing within the PN H$\alpha$ image are
approximately 1/500 (6~cm) and 1/200 (20~cm).  There are several other arguments that
imply a physical linkage with the PN: they have statistically the same nonthermal radio spectral 
indices instead of the more common thermal sources found along the Galactic plane; there may be a
faint connection between the emission of A and B; and A lies very close to the edge of the PN 
and is clearly curved, roughly concentrically, with the rim (Fig.~\ref{curve}), making a physical 
interaction between A and the ejecta that define the PN rim highly plausible.

\subsection{Radio luminosity and kinetic energy: comparison with WR stars}
We have compared our continuum observations at different epochs and see no indications of
variability at any of the wavelengths observed over a period of 1.5~yr.  
Chapman et al. (1999: their Table~3) tabulated the monochromatic 6-cm luminosity of WRs 
that emit nonthermal radio emission.  The average of the 9 WRs detected was 
2.1$\pm$0.6$\times$10$^{19}$~ergs~s$^{-1}$~Hz$^{-1}$. The same calculation yields 3.7 
and 0.7$\times$10$^{19}$~ergs~s$^{-1}$~Hz$^{-1}$ for A and B, respectively. Summing the two 
nonthermal sources in OH\,354.88-0.54, the PN is more luminous than the average nonthermal 
WR emitter. Adding A and B we find that the approximate nonthermal radio luminosity (for an adopted
bandwidth of 10~GHz; cf. Chapman et al. 1999) is 4.4$\times$10$^{29}$~ergs~s$^{-1}$~Hz$^{-1}$.
High energy emission is detected from some WRs in which colliding winds occur.  OH\,354.88-0.54
has not been detected based on a search of X-ray and $\gamma$-ray catalogues.  In particular, 
Chandra has not observed in its direction.

We now compare OH\,354.88-0.54 with WR stars, for which a typical radio luminosity is a 
few $\times10^{29}$~ergs~s$^{-1}$.  The total kinetic energy (KE) flux calculated from the 
known mass loss rates (\.{M}) of WRs (a few $\times$10$^{-5}$~M$_\odot$~yr$^{-1}$) and wind velocities
(V$_{wind}$$\sim$2000~km~s$^{-1}$), $\frac{1}{2}$\,\.{M}~V$_{wind}$$^2$,
 is typically 6$\times$10$^{37}$~ergs~s$^{-1}$. For a WR$-$O-star collision about 0.3\% 
of the flux passes through the interaction region, so the KE flux
at the interaction is about 2$\times$10$^{35}$~ergs~s$^{-1}$ (see Chapman et al. (1999)
for details).

For the detected continuum sources A and B, the 5-GHz radio luminosity is 
4.4$\times$10$^{19}$~ergs~s$^{-1}$~Hz$^{-1}$
and the total integrated radio luminosity (L$_{radio}$) is thus about 4.4$\times$10$^{29}$~ergs~s$^{-1}$.
We assume that the ratio of KE flux/radio luminosity is the same as for WR stars (about 
10$^6$:1).  Thus approximately 4.4$\times$10$^{35}$~ergs~s$^{-1}$ should be available at the shock 
regions.  By comparison with WR binaries, the total KE flux of the fast wind of OH\,354.88-0.54 is not 
well determined as we do not know
what fraction of the fast wind encounters the shocked regions.  Source A occupies about 1/100
of the total shell area (estimated at 6~cm) while source B is smaller.  For a smooth distribution 
this suggests that the KE could be about 100$\times$ higher than at the shock.  However, if
the fast wind is clumpy or asymmetric then this is likely to be considerably overestimated.

As an example, we consider a scenario where 10\% of the total KE flux reaches the detected shocked regions 
giving a total KE flux of 4.4$\times$10$^{36}$~ergs~s$^{-1}$.  For V$_{fast}$, the velocity of the 
fast wind, we adopt 1000~km~s$^{-1}$.  From these values we can crudely estimate the mass-loss rate
of the fast wind to be \.{M}$_{fast}$$\approx$1.2$\times$10$^{-5}$~M$_\odot$~yr$^{-1}$.

We now estimate the velocity of the compressed shell (V$_{shell}$) for this mass-loss rate.
From eqn.(9) of Zijlstra et al. (2001) and using \.{M}$_{agb}$~=~7$\times$10$^{-5}$, V$_{agb}$~=~15~km~s$^{-1}$, 
\.{M}$_{fast}$~=~1.2$\times$10$^{-5}$, and V$_{fast}$~=~1000~km~s$^{-1}$,
we have $\mu$~=~0.17 and $\xi$~=~67, where $\mu$~=~\.{M}$_{fast}$/ \.{M}$_{agb}$
and $\xi$~=~V$_{fast}$/V$_{agb}$.  For these values the velocity of the compressed shell is $\sim$60~km~s$^{-1}$
and this can be compared to the observed line-of-sight optical velocity of 20~km~s$^{-1}$.  
If less than 10\% of the total
energy flux reaches the shocked regions then both \.{M}$_{fast}$ and V$_{shell}$ would be higher while, 
if a larger fraction of the energy flux were intercepted, then \.{M}$_{fast}$ and V$_{shell}$ would be lower.

\subsection{Wind interactions}
Despite the roughness of these estimates they do provide an idea of the relevant
interactions.  The fast wind ploughs into the slow AGB wind creating shocks,
non-thermal emission and a compressed shell on the inner face. If the
fast wind has a momentum (\.{M}$_{fast}$$\times$V$_{fast}$) that is roughly comparable to (or
greater than) that of the slow wind then, when they collide, the slow wind is
compressed and accelerated and this increase in velocity is indicated by
the velocity of the optical emission.  The shock excitation must be stronger in the 
compressed intranebular shell than at the ISM and the rather high density derived
from the red [S{\sc ii}] doublet is more suggestive of a post-shocked region
rather than a typical PN envelope.  Therefore, it is
more likely that the optical emission in the PN arises at the interface of the fast 
wind and AGB wind, rather than at the interface of the AGB wind and the ISM.  

The geometry of the emerging fast wind is unknown because any
spherically symmetric flow from the star would be immediately modified by its encounter with the dense
circumstellar OH toroid.  Naively, were this toroid to constrain the fast wind to the polar directions
(north-south) in which OH maser emission is weakest (Welty, Fix \& Mutel 1987), one might expect to see
nonthermal ``hot spots'' in two opposing directions.  There is evidence of a 10-$\mu$m source around
the star that is significantly elongated north-south (Cobb \& Fix 1987) but this need not trace any
dynamical structure.  In fact, Figure~\ref{quotpsf} suggests that sources A and B do not correspond to 
a bipolar wind encountering a spherical distribution of slowly-moving ejecta.  Consequently, one might 
have to invoke either a fast wind capable of breaking out of the OH toroid in any region of locally reduced
density, or a non-spherical distribution of the original PN ejecta, or both.  However, Table~2 shows 
that both sources appear projected in the sky at the same radial distance from the central star.
This is consistent with a spherical ejection for the slow wind.  The natural direction for
any asymmetry of the fast wind or of the original debris would be in the direction roughly from northwest to
southeast of the star, based on the morphology of the H$\alpha$ nebular brightness and the line
bisecting sources A and B through the possible weak bridge of emission observed at 13~cm (Fig.~\ref{13cm}).
There is also evidence for a preferred axis in the same p.a. in the OH velocity slices between $-5$ and
$-8$~km~s$^{-1}$ published by Welty et al. (1987: their Fig.~15).  At more negative velocities the OH
brightness indicates weaker OH emission along a north-south axis.  None of these directions appears
to be associated with the nonthermal shock zones.

Strong nonthermal radio continuum emission has also been observed from the OH/IR star OH\,326.53-00.419 
(also known as D046), with a spectral index of $-0.8$.  The OH maser emission from D046 occurs over 
an unusually broad range of 80~km~s$^{-1}$ (Sevenster \& Chapman 1999; Sevenster \& Chapman 2005).
D046 appears to represent a post-AGB object, in transition between the AGB
and PN phases. These nonthermal emitters may all arise because of the impact of fast winds on 
circumstellar debris near the end of AGB life.  OH\,354.88-0.54's intermediate mass progenitor 
may confer its uniqueness, in that it is still a pulsating long-period variable after it has 
already created a visible PN, simply because the time scale for stellar evolution is much shorter
for a 4~M$_\odot$ star than for a solar-mass object.

\section{Conclusions}
The ionization of the PN associated with OH\,354.88-0.54 appears to arise through
shock interactions between the AGB star's fast wind and previously ejected cold nebular
material.  Our discovery of patches of nonthermal radio emission in this PN suggests
a kinship with colliding-wind WR binaries but, in OH\,354.88-0.54, the two winds
were produced by a single central star.  The locations of the shocked, nonthermally 
emitting, regions do not suggest a spherical interaction zone between the winds, nor do
they align with any geometrically meaningful direction in the PN itself.  However, the 
nonthermal regions each show the same projected distance from the central star in the 
plane of the sky, and there is the possibility that faint radio emission links the two
regions.  These characteristics suggest that the anisotropic fast wind collides with the
continuous inner spherical surface of the slow wind.
Therefore, this PN offers a rare opportunity to recognize the nonspherical geometry of the 
fast wind, close to the time of its onset.  

\section{Acknowledgments}
The Australia Telescope is funded by the commonwealth of Australia for operation
as a National Facility managed by the CSIRO.  The MOST is owned and operated by
the University of Sydney, with support from the Australian Research Council and
Science Foundation within the School of Physics.
MC thanks NASA for supporting this work under its Long Term Space Astrophysics and Astrophysics
Data Analysis programmes, through grants NAG5-7936 and NNG04GD43G with UC Berkeley.
The AAO has undertaken the SHS on behalf of the astronomical community. We thank Jim Caswell, David Frew
and Lister Staveley-Smith for valuable discussions.\\

{}
\end{document}